\documentclass{PoS}
\usepackage{amssymb,amsmath}
\usepackage{cancel}

\graphicspath{{Figures/}}

\title{A local fermion update algorithm for supersymmetric Yang-Mills quantum mechanics}

\ShortTitle{A local fermion update algorithm for SYM quantum mechanics}

\author{Georg Bergner, Hang Liu and \speaker{Urs Wenger} \\
        Albert Einstein Center for Fundamental Physics\\
         Institute for Theoretical Physics\\
        University of Bern\\
        Sidlerstrasse 5\\
        CH--3012 Bern\\
        Switzerland\\
        E-mail:  \email{bergner@itp.unibe.ch}, \email{liu@itp.unibe.ch}, \email{wenger@itp.unibe.ch}}

\abstract{ 
  We present a local fermion update algorithm for ${\cal N}=4$
  supersymmetric Yang-Mills quantum mechanics which allows simulations
  in fixed canonical sectors.  We discuss some aspects of the physics
  of this theory, including the appearance of flat directions in the
  bosonic potential and the metastabilities related to those. In
  particular, we show that the modulus of the bosonic fields diverges
  in some of the fermion sectors and for thermal boundary conditions.
}

\FullConference{34th annual International Symposium on Lattice Field
  Theory\\
 24-30 July 2016\\
University of Southampton, UK}

\begin{document}

\section{Introduction}
These proceedings report on our ongoing effort to simulate
supersymmetric Yang-Mills (SYM) gauge theories with gauge group
SU($N$).  Our considerations start from the dimensional reduction of
$\mathcal{N}=1$ SYM in $d=4$ dimensions with gauge group SU($N$) which
yields ${\cal N}=4$ SYM in one dimension, i.e. SYM quantum
mechanics. In the process the three spatial components of the
4-dimensional gauge field become bosonic fields denoted by $X_i(t), \,
i=1,2,3$, while the temporal component is denoted by $A(t)$. The
fermionic degrees of freedom are represented by complex 2-component
Grassmann fields $\overline{\psi}(t),\, \psi(t)$. All fields are in
the adjoint representation and the complete action reads
\cite{Catterall:2007fp,Catterall:2008yz}
\begin{align*}
  S=\frac{1}{g^2}\int_{0}^{\beta} dt \, \text{Tr} \left\lbrace  \left(  D_t X_i   \right)^2 -\frac{1}{2} \left[  X_i , X_j  \right]^2 
    +\overline{\psi} D_t \psi -\overline{\psi} \sigma_i \left[ X_i , \psi \right]
  \right\rbrace 
\end{align*}
where $D_t=\partial_t-i[A(t),\, \cdot\,]$ is the covariant derivative
and $\sigma_i$ are the Pauli matrices. The temporal extent $\beta$ of
the system is discretized in the Euclidean direction using $L_t$
lattice points. In order to maintain gauge invariance of the
discretized system, the continuum gauge field is replaced by gauge
links $U(t)$ which are elements of the gauge group SU($N$), and hence
the lattice covariant derivative becomes ${\hat D}_t X_i(t)= U(t)
X_i(t+1)U^\dagger(t)-X_i(t)$. The same difference operator also
emerges for the fermions after inclusion of the Wilson term which
breaks time reversal and hence charge conjugation
symmetry. Supersymmetry is broken by the discretization as
well. However, all symmetries are automatically restored in the
continuum limit.  For further reference, we include a chemical
potential, which couples to the fermion number, and write out the
fermion action as
\begin{align*}
{S}_F=\frac{1}{2g^2} \sum_{t=0}^{L_t-1} \,
\left[-\overline{\psi}_\alpha^a(t) W^{ab}_{\alpha\beta}(t) \, e^{\mu} \, \psi_\beta^b(t+1) +\overline{\psi}_\alpha^a(t)\Phi_{\alpha \beta}^{ab}(t)\psi_\beta^b(t) 
 \right] \, .
\end{align*}
The matrix $W$ connecting the fermion fields at different lattice
sites reads
\begin{align*}
W^{ab}_{\alpha\beta}(t) = 2 \delta_{\alpha\beta} \cdot\text{Tr}\{T^a U(t) T^b
U^\dagger(t)\} \, ,
\end{align*}
where $T^a$ are the generators of the gauge group SU($N$), and $\Phi$
is the $2 (N^2-1) \times 2 (N^2-1)$ Yukawa interaction matrix
\begin{align*}
\Phi_{\alpha \beta}^{ab}(t) = (\sigma_0)_{\alpha\beta} \cdot \delta^{ab} -
 2 \, (\sigma_i)_{\alpha\beta} \cdot \text{Tr}\{T^a [X_i(t), T^b]\} \, . 
\end{align*}
Here, $\sigma_0$ is the $2\times 2$ unit matrix.

\section{Canonical formulation}
Usually, simulations are done using the grand canonical partition
function
\begin{align*}
Z_{p,a} = \int{\cal D}U \, {\cal D}X_i \,  e^{-S_B[U,X_i]} \det {\cal
   D}_{p,a}[U,X_i;\mu] \, 
\end{align*}
where $S_B$ is the bosonic action and ${\cal D}_{p,a}$ the Wilson
Dirac operator with periodic or antiperiodic fermion boundary
conditions, respectively. Instead of fixing the chemical potential,
one can also fix the fermion number $n_f$ and work with the canonical
partition functions
\begin{align*}
Z_{n_f} &= \int{\cal D}U \, {\cal D}X_i \, e^{-S_B[U,X_i]} \det{\cal
  D}_{n_f}[U,X_i] \, .
\end{align*}
While the fugacity expansion relates the canonical and grand canonical
partition functions and provides a relation between the corresponding
determinants, the direct calculation of $\det{\cal D}_{n_f}$
constitutes a challenge for doing simulations directly in the
canonical formulation. It turns out that the temporal reduction of the
Wilson Dirac operator \cite{Steinhauer:2014oda,Alexandru:2010yb} based
on Schur complement techniques is a crucial step for performing this
task. The reduced determinant formula reads
\begin{align*}
\det {\cal D}_{p,a}[U,X_i; \mu] = \det \left[ {\cal T} \mp e^{+ \mu L_t} \right]\;, \quad
{\cal T} = \prod_{t=0}^{L_t-1} (\Phi(t)  W(t)) \, 
\end{align*}
where the matrix ${\cal T}$ is of size $n_f^\text{max} \times
n_f^\text{max}$ with $n_f^\text{max} = 2 (N^2-1)$. Comparing this
expression with the fugacity expansion and using some algebraic matrix
identitites one obtains an explicit formula for the canonical
determinants in terms of transfer matrices defined for fixed canonical
sectors \cite{Steinhauer:2014oda},
\begin{align}
\label{eq:principal minors}
\det {\cal D}_{n_f}[U,X_i]  = \text{Tr} \prod_{t=0}^{L_t-1}
\left[T_{n_f}^\Phi(t) \cdot T^W_{n_f}(t)\right] = \text{Tr}
M_{n_f}({\cal T}) = \sum_B \det{\cal T}^{\, \bcancel{B} \, \bcancel{B}}[U,X_i] 
\, .
\end{align}
Here, $T_{n_f}^\Phi(t) \cdot T^W_{n_f}(t)$ constitutes the transfer
matrix at time $t$, which describes the transition probabilities of
the $n=\left( \genfrac{}{}{0pt}{}{n_f^\text{max}}{{n_f}}\right)$ states in sector
$n_f$, and $M_{n_f}(\cdot)$ denotes the matrix of minors of rank
$n_f$, i.e., the index set $B \subseteq \{1,2,\ldots,2(N^2-1)\}$ is of
size $n_f$ \cite{Steinhauer:2014oda}.

\section{Simulation algorithms}
There are several possibilitites to simulate the canonical partition
functions based on eq.(\ref{eq:principal minors}). Results from
simulating directly the transfer matrices will be described elsewhere,
while here we concentrate on a simulation strategy based on the sum of
the principal minors,
\begin{align*}
Z_{n_f} = \sum_B \int{\cal D}U \, {\cal D}X_i \, e^{-S[U,X_i]} \det {\cal
  T}^{\bcancel{B}\bcancel{B}}[U,X_i] \, .
\end{align*}
Since the number $n$ of principal minors of order $n_f\sim
n_f^\text{max}/2$ grows factorially with the size $n_f^\text{max}$ of
${\cal T}$, an exact evaluation of the canonical determinant is
impractical or even impossible. Instead, we employ an efficient
stochastic evaluation of the sum of the principal minors and treat the
summation index set $B$ as an additional degree of freedom of the
system. Thereby, the index set $B$ is dynamically updated using a
standard Metropolis algorithm. Starting from the index set $B$ a new
random set $B'$ is proposed using Fisher-Yates reshuffling and the
transition $B \rightarrow B'$ is accepted with the probability $p_{B
  \rightarrow B'} = \min[1,A_{B \rightarrow B'}]$ with
\begin{align*}
A_{B \rightarrow B'} = \left|
\frac
{\det {\cal T}[U,X_i]^{\bcancel{B}'\bcancel{B}'}}
{\det {\cal T}[U,X_i]^{\bcancel{B}\bcancel{B}}}
\right| \, .
\end{align*}
Then, the remaining fields are updated keeping the index set fixed. In
order for this whole process to be practical, one needs an efficient
calculation, or update, of the full matrix ${\cal T}$. We are using a
binary tree data structure to store intermediate products, such that
only ${O}(\ln L_t)$ matrix multiplications are necessary instead of
${O}(L_t)$.

\begin{figure}[t!]
\centering
\includegraphics[width=0.49\textwidth]{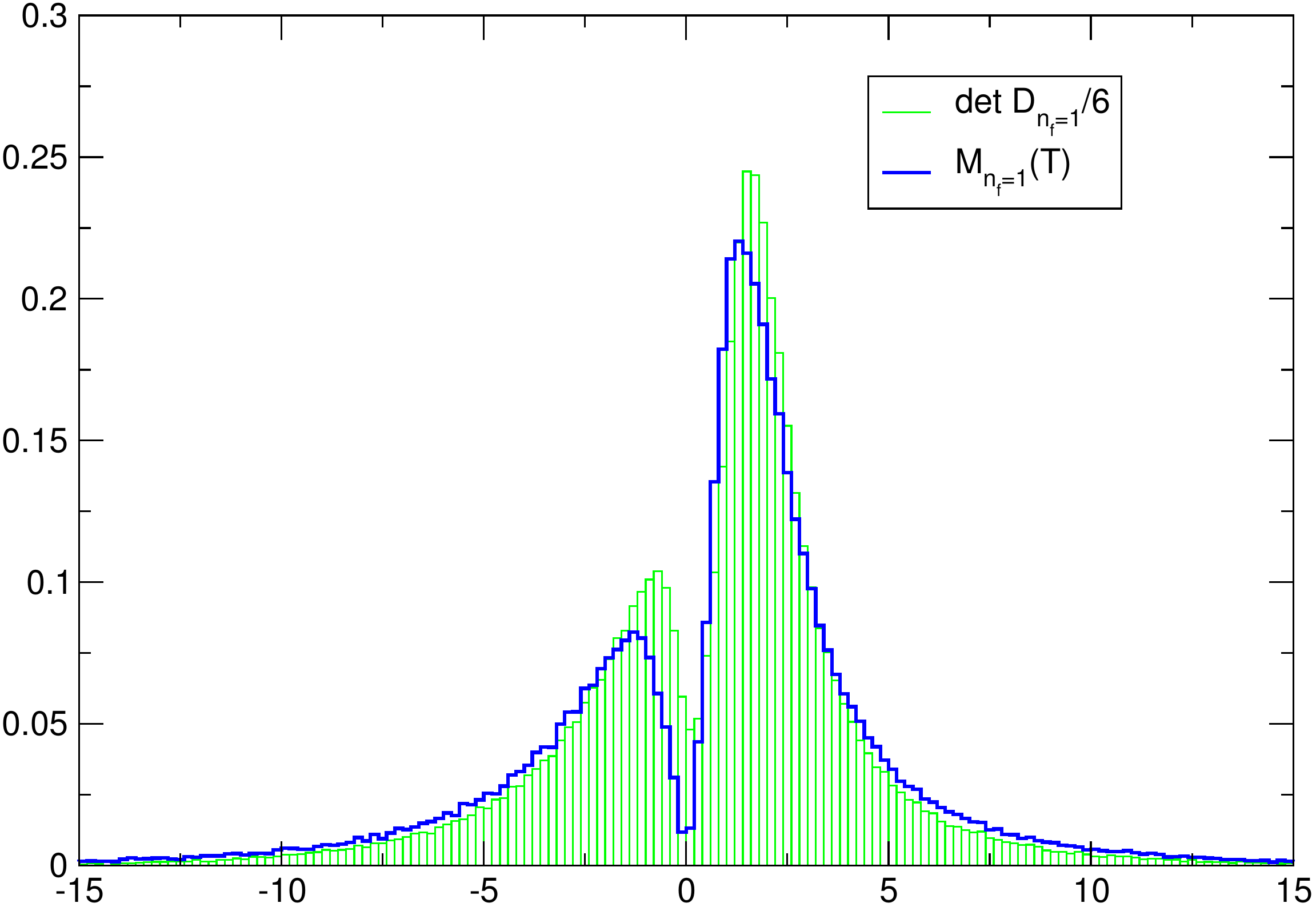}
\hfill \includegraphics[width=0.49\textwidth]{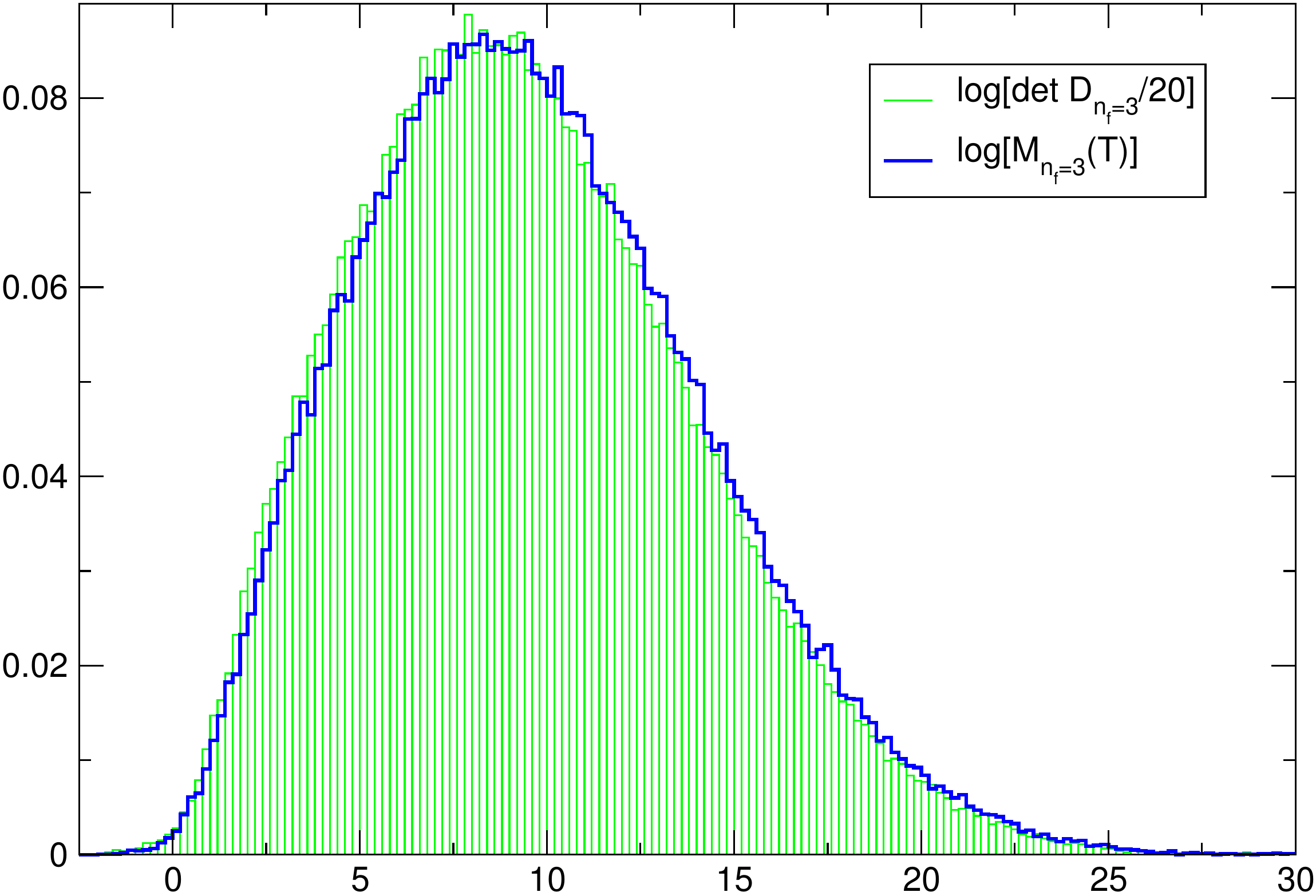} 
\caption{Distributions of the principal minors $M_{n_f}({\cal T})$
  from the stochastic evaluation of the trace
and the corresponding canonical determinant for SU(2), $L_t=24$ at
$\beta=1.2$ in sector $n_f=1$ (left plot) and $n_f=3$ (right plot).
\label{fig:minor_distributions}
}
\end{figure}

The prinicipal minors need not be positive, but can have negative
signs, in which case reweighting with the reweighting factors $R_{n_f
} =\text{sign}(M_{n_f})$ is required.\footnote{Reweighting between
  different fermion sectors is also possible and turns out to be
  reliable in certain cases, cf.~Ref.~\cite{Bergner:2015ywa}.}  As
illustrated in the left plot of Fig.~\ref{fig:minor_distributions},
the Metropolis algorithm has no difficulty to tunnel between the
sectors of the configuration space which yield negative and positive
contributions, but in fact samples both contributions very
efficiently. This is due to the fact that the Metropolis algorithm
allows the proposal and occasional acceptance of large changes of the
fields on the one hand, and sign changes in the determinant from
updating the index set on the other hand. We further note that the
negative contributions are absolutely necessary for reweighting the
configurations to different fermion number sectors, but they are not
frequent enough to generate a severe sign problem.  In fact, negative
contributions do not occur in all sectors. For SU(2) for example,
contributions in sectors $n_f=0, 6$ are strictly positive, while
sectors $n_f=1,5$ have significant negative contributions. In sectors
$n_f=2,3,4$ the negative contributions are negligible, cf.~right plot
of Fig.~\ref{fig:minor_distributions}.

Also shown in Fig.~\ref{fig:minor_distributions} are the distributions
of the principal minors $M_{n_f}({\cal T}) = \det {\cal
  T}[U,X]^{\bcancel{B}\bcancel{B}}$ as they occur in the stochastic
evaluation of the sum, i.e.~the trace $\text{Tr} M_{n_f}$, in comparison
with the distribution of the corresponding full canonical determinant
divided by the number of principal minors in the given sector. We see
that the principal minors follow the full determinant very closely and
evolve the system very similarly, but their evaluation is by a factor
$n$ faster. While for SU(2) this factor is maximally 20, for SU(3) the
maximal gain is already 12870.

\begin{figure}[t!]
\centering
\includegraphics[height=0.36\textwidth]{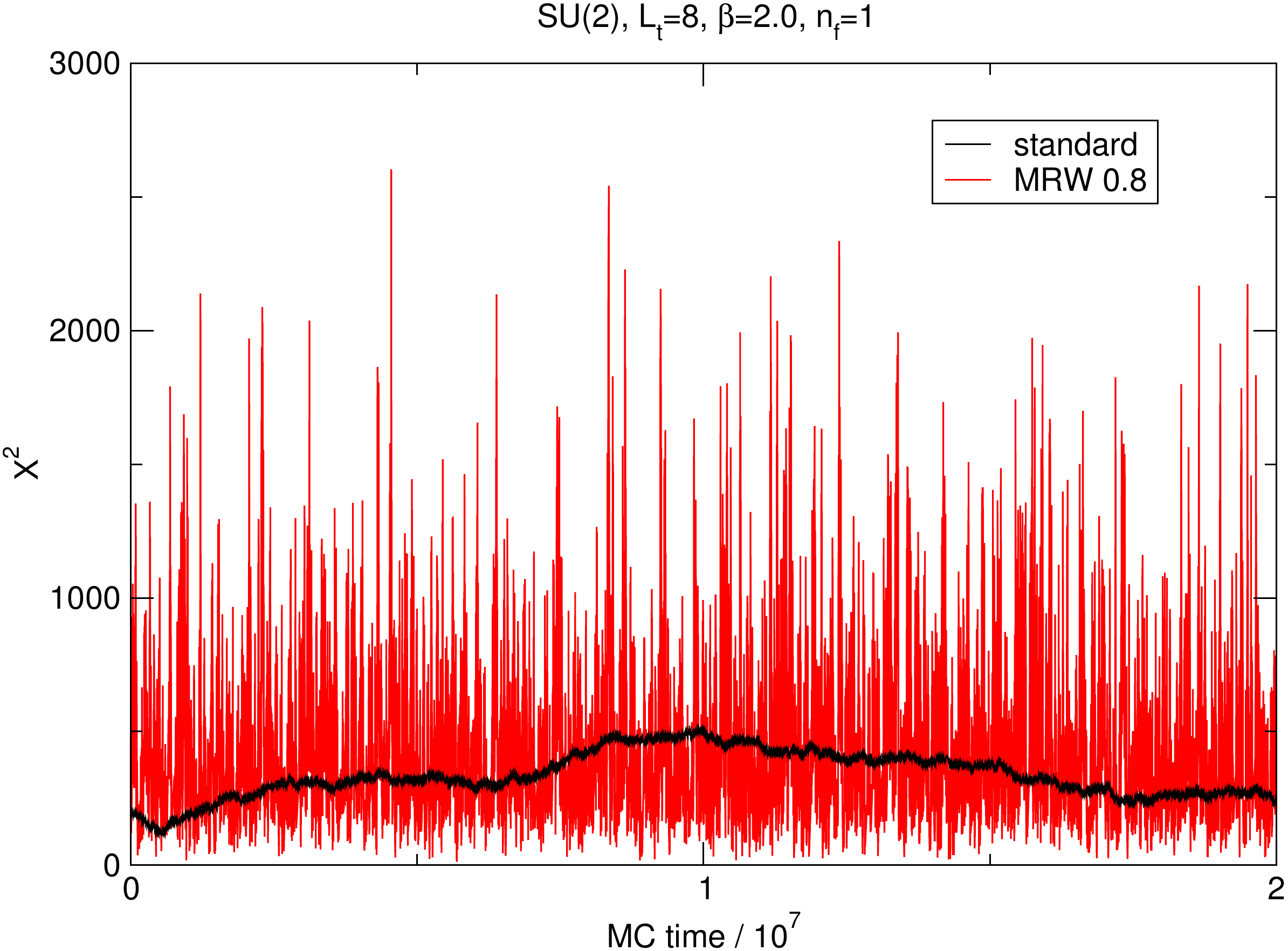} \hfill \includegraphics[height=0.36\textwidth]{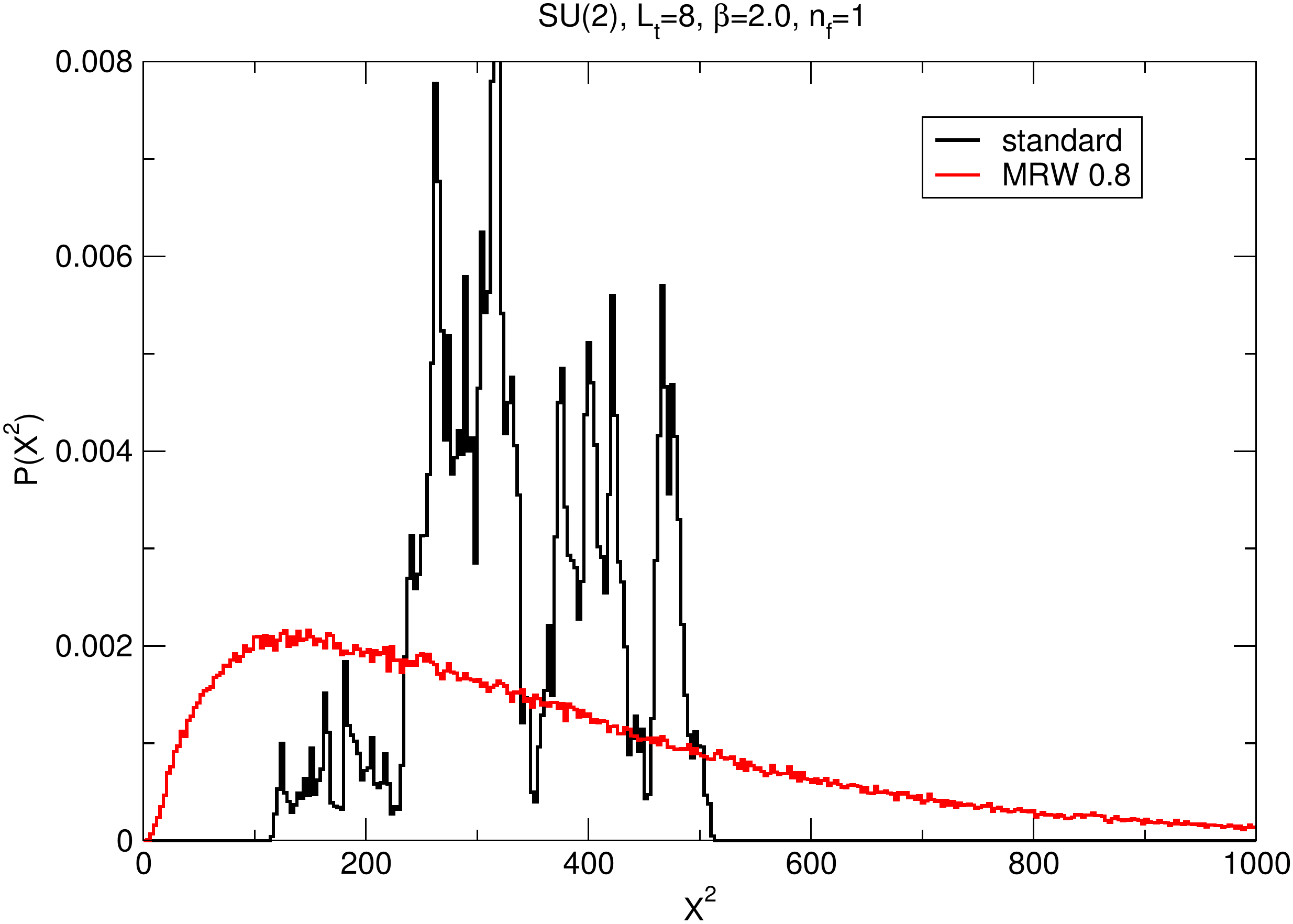} 
\vspace{-0.5cm}
\caption{
MC time history (left plot) and the corresponding
distribution (right plot) of the moduli $X^2$ for  SU(2) on a $L_t=8$
lattice at $\beta=2.0$ in the $n_f=1$ sector, once with the standard
Metropolis algorithm and once with the multiplicative random walk
(MRW) algorithm. \label{fig:multiplicative random walk}}
\end{figure}

One peculiarity of the model considered here is the fact that it
possesses so-called flat directions (moduli space) along which $\left[
  X_i , X_j \right] = 0$. As a consequence, the system may suffer from
$X_i$ running away and $X^2 \equiv1/L_t \sum_{t=0}^{L_t-1}\text{Tr}\{ X_i(t)
X_i(t) \}$ becoming arbitrarily large. While one can regularize the
divergence with the deformation $m^2 X^2$, we observe that metastable
states along the flat directions may be introduced due to an interplay
between lattice artefacts and the deformation, and the simulation
consequently suffers from critical slowing down.  Essentially, in
those metastable states fluctuations orthogonal to the flat direction
are highly suppressed compared to those along the flat direction,
hence the standard Metropolis becomes highly inefficient. A solution
to this problem is provided by the multiplicative random walk (MRW)
update algorithm which updates the bosonic fields $X$ {\it
  collectively} by rescaling them by a global random factor $R$. In
order to fulfill detailed balance, the acceptance probability has to
be chosen with care. The procedure is illustrated in
Fig.~\ref{fig:multiplicative random walk}
where we show the MC time history (left plot) and the corresponding
distribution (right plot) of the moduli $X^2$ for a system which is
stuck in a flat direction (for SU(2) on a $L_t=8$ lattice at
$\beta=2.0$ in the $n_f=1$ sector). With the standard Metropolis
algorithm $X^2$ moves very slowly and the huge autocorrelations are
reflected in the completely unreliable distribution of $X^2$. In
contrast, the MRW algorithm samples $X^2$ very efficiently and
produces a smooth distribution which in turn allows a reliable
extraction of $\langle X^2\rangle$.

\section{Results}
We first investigate the behaviour of the stable and metastable phases
\begin{figure}[t]
\centering
\includegraphics[width=0.49\textwidth]{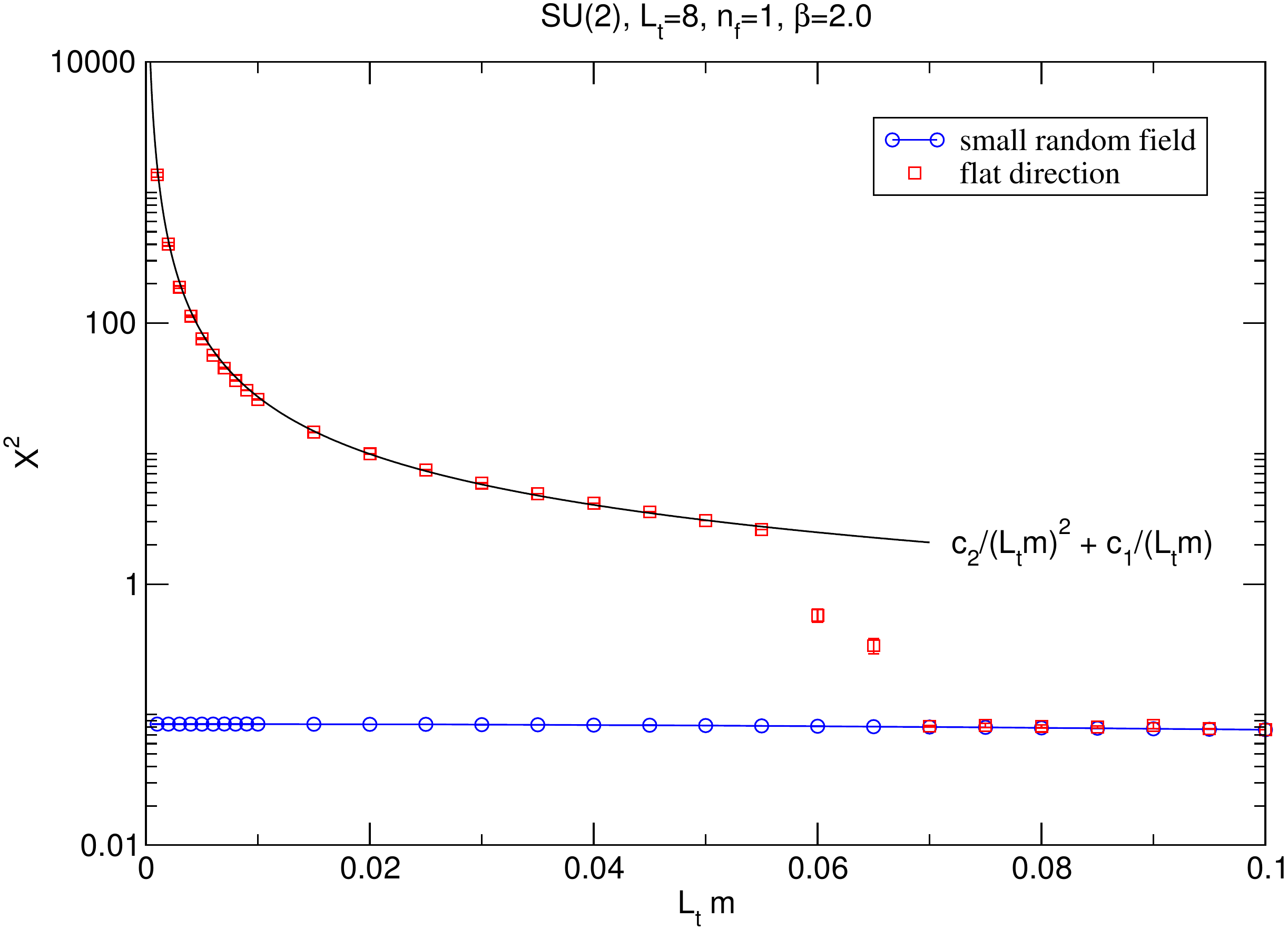}
\hfill
\includegraphics[width=0.49\textwidth]{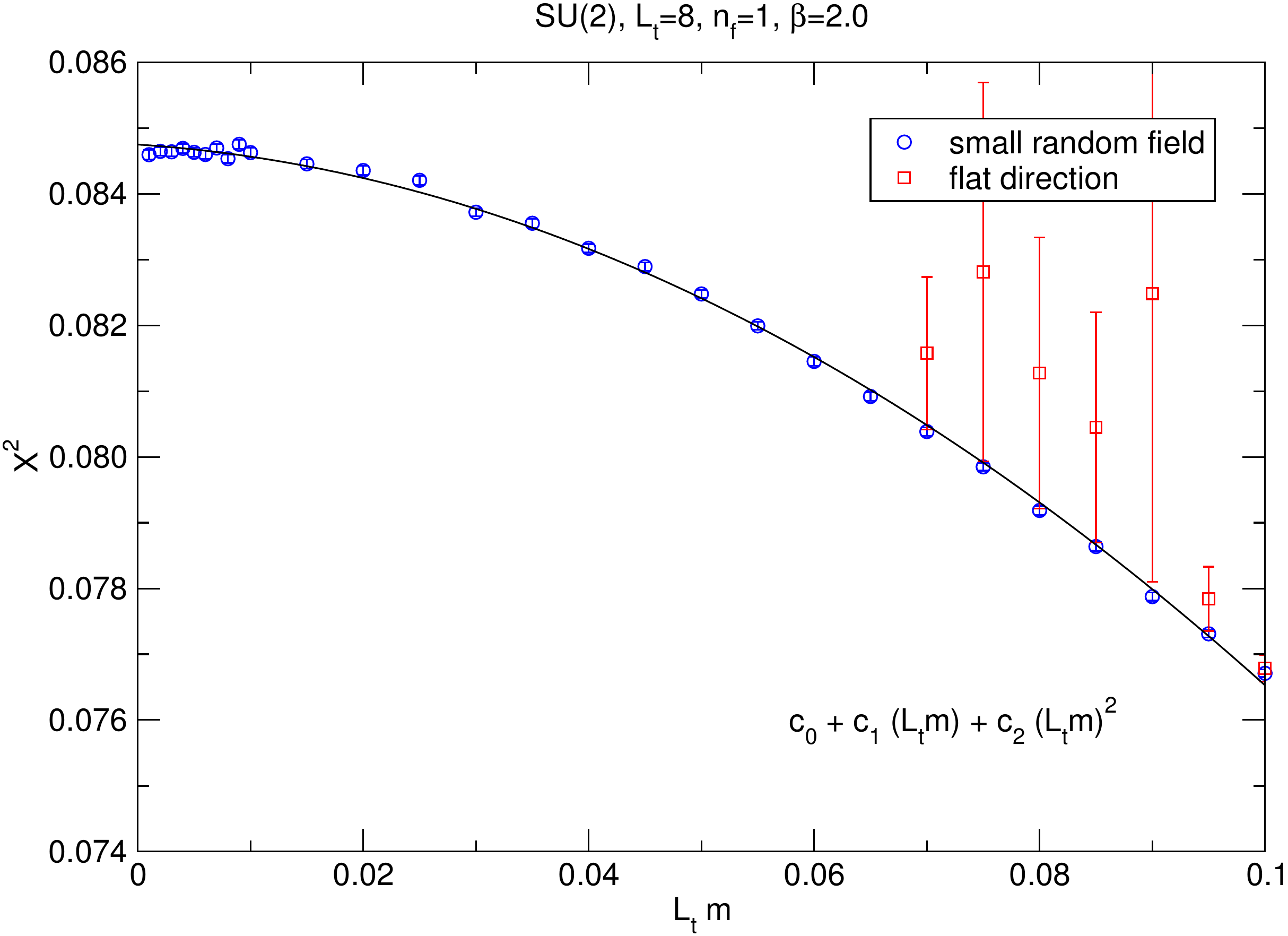}
\caption{The expectation value of the moduli $X^2$ as a function of
  the regulator mass for SU(2) on a $L_t=8$ lattice at $\beta=2.0$ in
  the $n_f=1$ sector, once starting from a configuration consisting of
  small random fluctuations, and once with a a large added component
  along a flat direction. The right plot is a zoom to the small region
  of $X^2$. \label{fig:metastable phase} }
\end{figure}
as a function of the regulator mass $m$ in the limit $m\rightarrow
0$. In order to do so, we prepare the system with a starting
configuration consisting of small fluctuations around $X=0$ (small
random field) on the one hand and with an added large component along
a flat direction on the other hand. As can be seen from
Fig.~\ref{fig:metastable phase}, for sufficiently large $m$ a system
prepared in a phase along the flat direction always tunnels back into
the phase exhibiting small fluctuations only, while for $L_t m
\lesssim 0.07$ the tunneling barrier is too large and hence the system
becomes metastable. The minimum of the metastable phase moves to
infinity and $\langle X^2\rangle$ diverges in the limit $m \rightarrow
0$ as $1/(L_tm)^2$ to leading order. Let us emphasize that the MRW
algorithm is crucial to obtain reliable results in the metastable
phase. In contrast, simulations starting from small random field
configurations seem stable and $\langle X^2\rangle$ is well behaved in
the limit $m \rightarrow 0$. In fact, for sufficiently large $L_t$
even simulations at $m=0$ are possible.

\begin{figure}[t]
\centering
\mbox{}\hspace{-0.5cm}\includegraphics[width=0.5\textwidth]{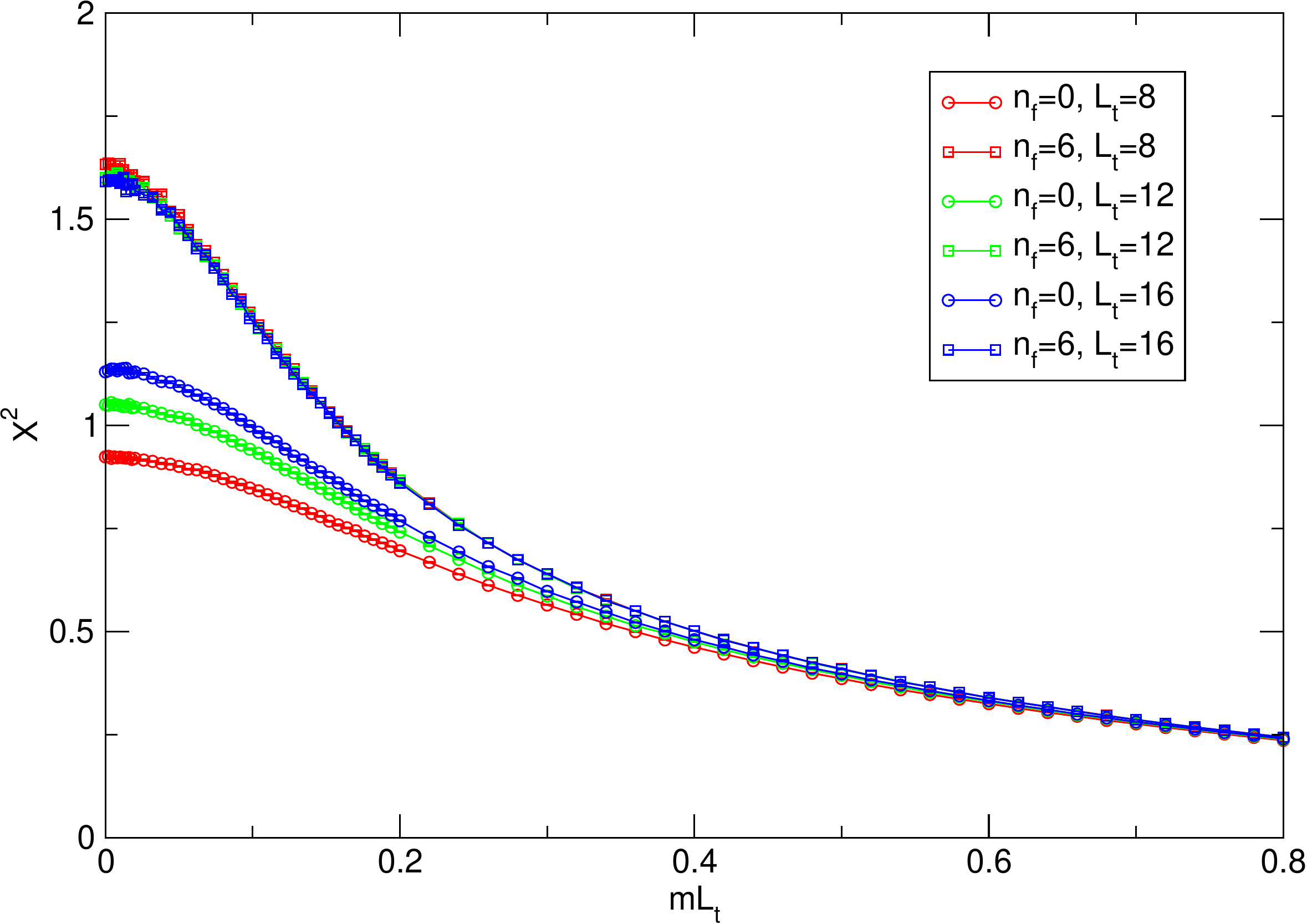}
\includegraphics[width=0.5\textwidth]{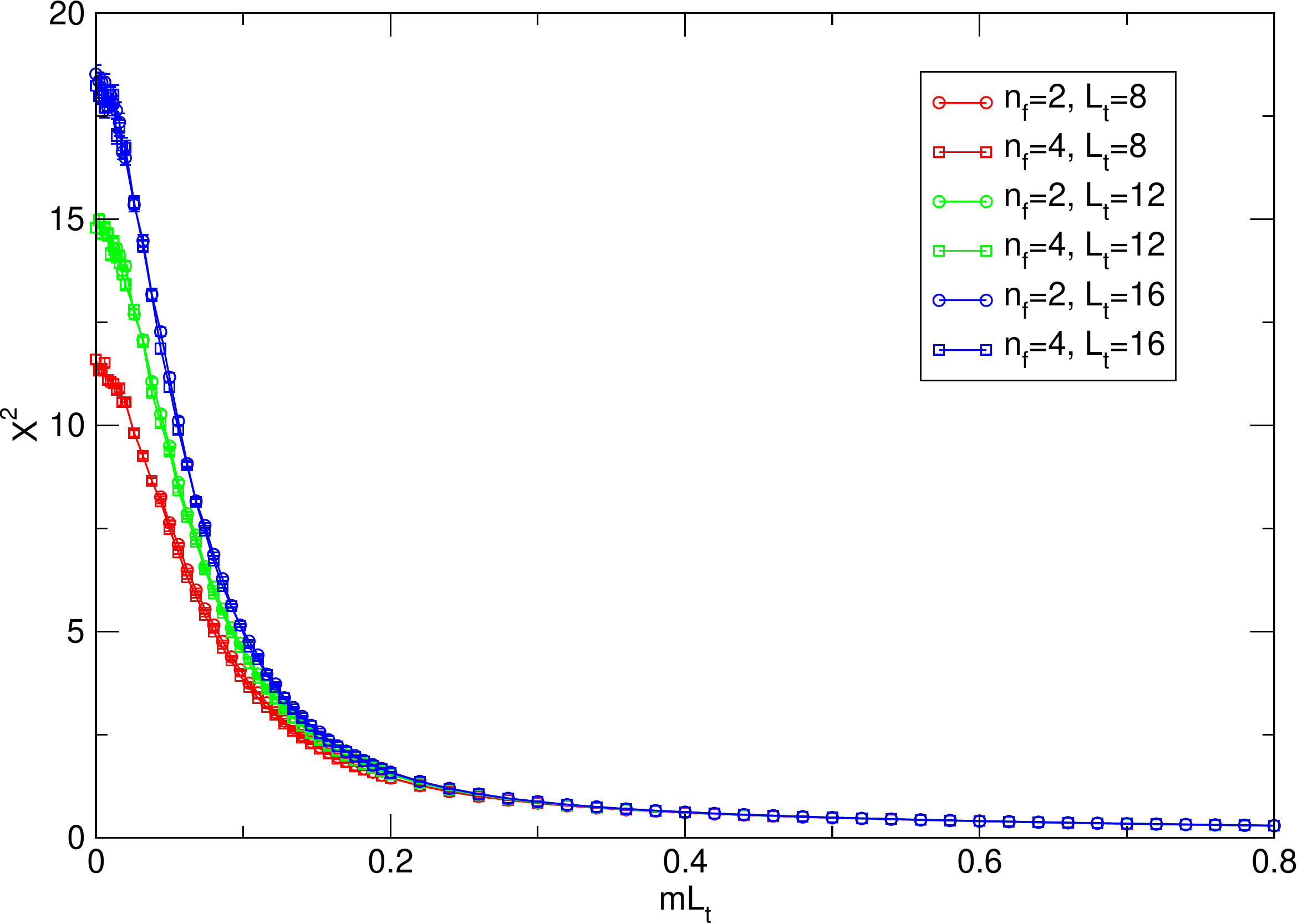}\\
\mbox{}\hspace{-0.5cm}\includegraphics[width=0.5\textwidth]{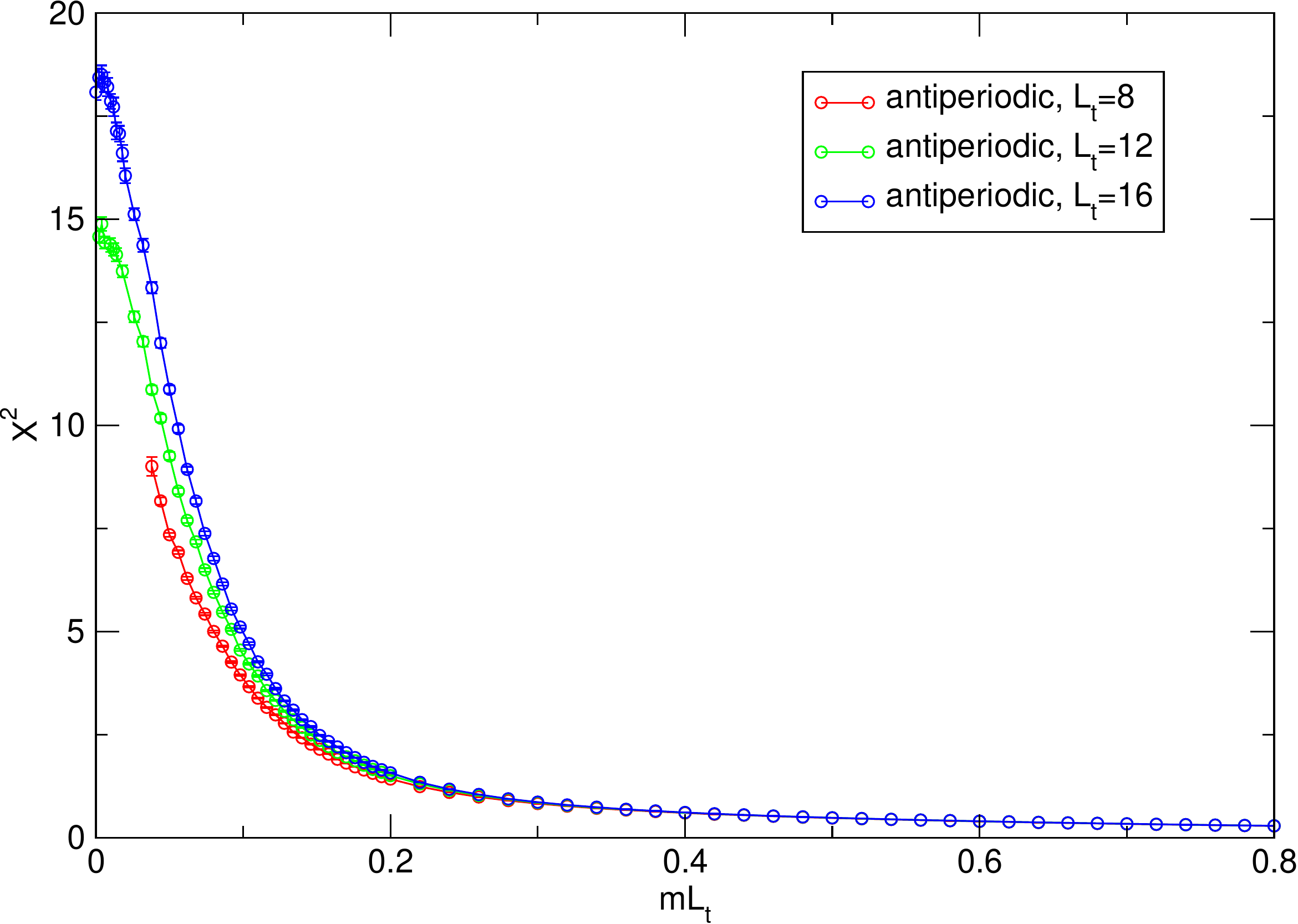}
\includegraphics[width=0.5\textwidth]{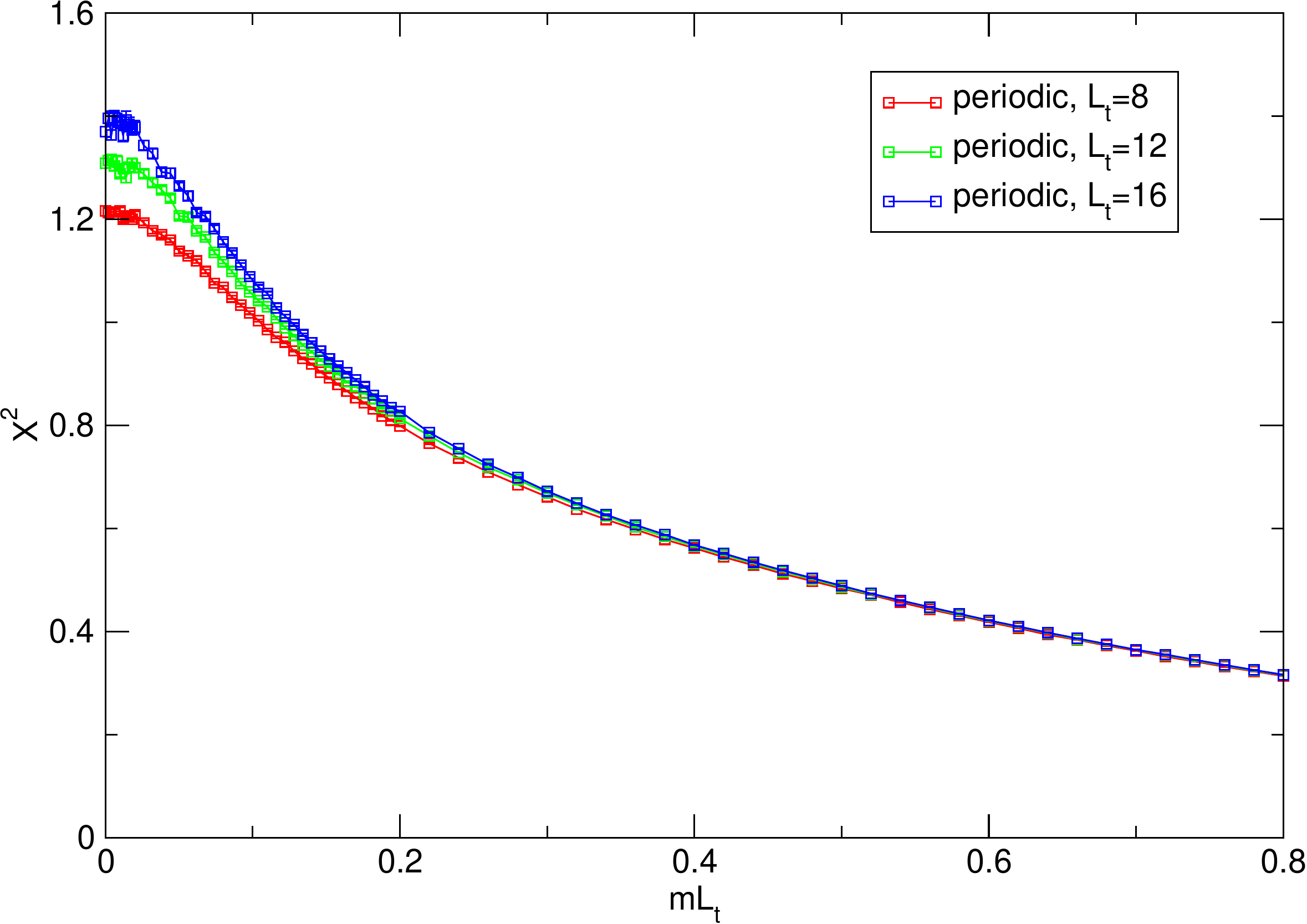}
\caption{The expectation value of the moduli $X^2 = \langle \text{Tr}
  X_i^2\rangle$ as a function of the regulator mass for SU(2) and
  various lattice extents $L_t=8, 12, 16$ at $\beta=0.5$ in the
  sectors with $n_f=0,\ldots, 6$ and for periodic and antiperiodic
  boundary conditions. \label{fig:X2_vs_mL} }
\end{figure}

Having clarified the role and fate of the metastable phase, we can now
investigate in detail the regulator dependence of the persisting
stable phase involving small fluctuations of $X$. In
Fig.~\ref{fig:X2_vs_mL} we show the results for $\langle X^2\rangle$
as a function of $m$ from simulations of SU(2) at $\beta=0.5$ for a
range of lattice extents $L_t$ in some
fermion number sectors, as well as for periodic and antiperiodic
fermionic boundary conditions. Here, $\langle X^2\rangle$ is rescaled
by $L_t$, and the continuum limit corresponds to $L_t\rightarrow
\infty$ at fixed $L_t m$.  Let us first discuss the lattice
artefacts. For large $m$ we observe universal scaling behaviour
independent of the lattice spacing $a=\beta/L_t$. In this regime,
lattice artefacts become very small and the degeneracy between the
charge conjugated fermion number sectors is restored. When the
regulator is removed, lattice artefacts become large and the effect
from the explicitly broken charge conjugation symmetry ${\cal C}$
becomes evident. Only for the sectors $n_f=2$ and 4 the ${\cal
  C}$-symmetry seems to remain exact at a finite lattice spacing. In
sector $n_f=6$ the lattice artefacts are tiny. The latter sector
corresponds to the quenched approximation, hence the leading artefacts
are expected to be $O(a^2)$ in contrast to the other sectors where
they are $O(a)$.

Next we discuss the behaviour of $\langle X^2\rangle$ at $m=0$ in the
continuum limit. First we note that simulations at $m=0$ at finite
$L_t$ seem to be possible - apparently, the simulations are stabilized
by lattice artefacts. Towards the continuum, the fermion sectors with
$n_f=2, 3, 4$~qualitatively show a different behaviour than the
sectors with $n_f=0, 1, 5, 6$. For the latter, $\langle X^2\rangle $
approaches a finite value in the continuum limit, while for the
former, $\langle X^2\rangle $ appears to diverge. Note that this
divergence has nothing to do with the divergence in the metastable
phase discussed before, but is expected due to zero energy states and
the spectrum becoming continuous in those sectors
\cite{Wosiek:2002nm,Campostrini:2004bs}. Similarly, $\langle X^2
\rangle$ is finite in the system with periodic b.c., because the
diverging contributions from sectors $n_f=2, 3, 4$ cancel each other,
while this does not happen for the system with antiperiodic b.c..

To summarize, we presented an efficient local fermion algorithm which
allows simulations in fixed canonical sectors in a completely
controlled way. Hence, our approach is an alternative to other
numerical efforts investigating supersymmetric Yang-Mills quantum
mechanics, such as \cite{Ambrozinski:2014oka,Berkowitz:2016jlq}.

\bibliography{alfuafSYMQM}

\end{document}